# Performance Tradeoffs of Joint Radar-Communication Networks

Ping Ren, Andrea Munari, Marina Petrova

*Abstract*— This letter considers a network where nodes share a wireless channel to work in turn as pulse radars for target detection and as transmitters for data exchange. Radar detection range and network throughput are studied using stochastic geometry tools. We derive closed-form expressions that identify the key tradeoffs between radar and communication operations. Results reveal interesting design hints, and stress a marked sensitivity of radar detection to communication interference.

## I. Introduction

Radar sensing functionalities are employed in conjunction with wireless data links to leverage environmental awareness in an increasing number of applications, ranging from indoor localization to automotive [1]. While radar and communication (comm) systems have been traditionally granted dedicated frequency channels [2], spectrum scarcity is triggering a rising interest into their coexistence over a shared band, e.g. in the 5GHz [3] or in the mm-wave domain [1]. A thorough understanding of the role played by mutual interference in possibly large networks of devices operating both in radar and comm mode is thus paramount for proper system design. From this standpoint, however, research efforts have focused so far mainly on toy topologies, studying interference mitigation techniques [3], [4] or evaluating the effect of radar interference on comm capacity [5]. Initial results in broader setups were offered in [6], proposing a Gaussian model for aggregate interference in an OFDM system, yet the fundamental tradeoffs of radar-comm coexistence remain largely unexplored.

To bridge this gap, we apply stochastic geometry tools to model a planar network of nodes that share a common wireless channel. Aiming at simple yet insightful results that capture the behavior of uncoordinated setups, we assume that each device operates alternatively as a pulse radar, attempting to detect a target, and as a comm transmitter, sending data packets following an ALOHA policy. Radar performance is studied in terms of maximum achievable detection range constrained to meeting a tolerable false alarm rate, while network throughput is employed to evaluate the efficiency of data exchanges. Leaning on some reasoned approximations, we derive for both metrics compact closed-form expressions that capture the effect of key parameters such as network density, antenna directionality, and time devoted to radar and comm operations. The derived formulations allow to explore the fundamental tradeoffs of the system, e.g. in terms of maximum achievable throughput while guaranteeing a desired detection range, and reveal non-trivial insights. Analytical results are verified by means of detailed network simulations.

P. Ren and A. Munari are with the Institute for Networked Systems of the RWTH Aachen University, D-52072 Aachen, Germany (e-mail: {firstname.lastname}@inets.rwth-aachen.de). M. Petrova is with the School of Electrical Engineering and Computer Science, KTH Royal Institute of Technology, 10044 Stockholm, Sweden (e-mail: petrovam@kth.se).

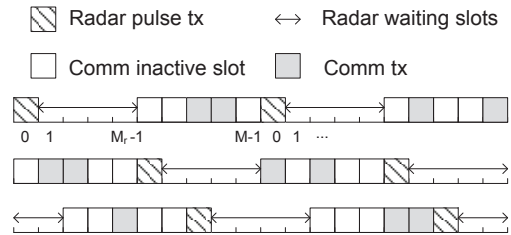

Fig. 1. Example timeline of three nodes over two radar-comm cycles.

## II. System Model

We focus on a set of wireless nodes, whose locations on the plane are modelled as a homogeneous Poisson point process (PPP) $\Phi = \{\mathbf{x}_i\}$ of intensity $\lambda$ over $\mathbb{R}^2$. Nodes share a common bandwidth, and each of them operates in turn in radar mode – trying to detect a target – and in comm mode – sending data packets to a desired receiver. More specifically, time is slotted, and a node follows a regular pattern of duration $M = M_r + M_c$ slots (or time units). Within the first $M_r$ time units, the node works as a pulse radar, transmitting a pulse and then waiting for a target echo for the subsequent $M_r - 1$ slots. The remaining $M_c$ time units are instead used for data delivery. Here nodes access the shared channel following a slotted ALOHA policy, transmitting a packet with probability $q_c$ over each slot towards a dedicated receiver, randomly located over a circle of radius $d_c$ centered at the transmitter. The uncoordinated nature of the network under study is captured allowing random offsets among the operating cycles of different nodes. Indeed, each node is assigned an independent mark $\nu_i$, uniformly chosen in the set $\{0, \ldots, M-1\}$, and transmits its radar pulses at time slots $\nu_i + nM$, $n \in \mathbb{N}$, as illustrated in the example timeline of Fig. 1. Throughout our discussion, we indicate as $\varepsilon \in [0, 1]$ the fraction of time devoted by a node to radar detection, i.e. $\varepsilon = M_r/M$, denoting the update rate of radar results. Transmissions both in radar and comm mode are performed with power $P_t$. Nodes share a common antenna pattern with gain $G = 4\pi/\varphi^2$ over the main beam of width $\varphi \in (0, 2\pi]$ and $0$ elsewhere, with bore-sight direction pointing towards the associated comm receiver. Signal propagation undergoes power-law path loss with exponent $\alpha > 2$, and noise is neglected given the interference-limited nature of the network.

To derive the performance of the system we focus, without loss of generality, on the typical node $\mathbf{x}_0$, located at the origin and transmitting radar pulses at slots $nM$, $n \in \mathbb{N}$ (i.e. with a slight abuse of notation, with mark $\nu_0 = 0$). When operating in radar mode, the device bases its detection decisions on incoming power. The target return signal power follows from the well-known radar equation as

$$\mathcal{S} = \frac{P_t G^2 c^2 \sigma}{(4\pi)^3 f^2} d_r^{-2\alpha} = K \frac{\sigma}{4\pi} d_r^{-2\alpha}$$

where $K := P_t(Gc/(4\pi f))^2$, $d_r$ is the target distance, $\sigma$ its radar cross section, $f$ the carrier frequency, and $c$ the speed of light. On the other hand, over a generic slot of interest $m$, the aggregate interference can be expressed as

$$\mathcal{I}_m = \sum_{i \in \Phi_r \cup \Phi_c/\{\mathbf{o}\}} K\|\mathbf{x}_i\|^{-\alpha}. \quad (1)$$

Into (1), $\Phi_r$ and $\Phi_c$ are independently thinned versions of $\Phi$, capturing the positions of nodes that transmit a radar pulse or a comm packet over slot $m$, respectively. Specifically, $\Phi_r$ and $\Phi_c$ are still homogeneous PPPs with intensity $\lambda_r = (\varepsilon/M_r)(\varphi/(2\pi))^2\lambda$ and $\lambda_c = (1-\varepsilon)q_c(\varphi/(2\pi))^2\lambda$. Here $(\varphi/(2\pi))^2$ accounts for the probability that the radiation pattern of a node overlaps with the one of the typical receiver, and $\varepsilon/M_r$ and $(1-\varepsilon)q_c$ for the fraction of slots a node transmits radar pulses and comm packets, respectively.

Following a power detection rule, a radar declares the presence of a target when the received power is above a predefined detection threshold $\theta$. Thus, a target is correctly detected over slot $m$ with probability $\mathsf{P}_\mathsf{d} = \mathbb{P}\{\mathcal{S} + \mathcal{I}_m > \theta\}$. Conversely, a false alarm occurs if the received power exceeds $\theta$ in at least one of the $M_r - 1$ slots in the absence of a target. The false alarm probability can then be expressed as

$$\mathsf{P}_\mathsf{f} = 1 - \mathbb{P}\{\mathcal{I}_1 \leq \theta, \mathcal{I}_2 \leq \theta, ..., \mathcal{I}_{M_r-1} \leq \theta\}. \quad (2)$$

Following common practice, we set the detection threshold so as to guarantee a target false alarm probability, and study the radar performance in terms of radar range.[1] In particular, we focus on the maximum distance $\mathsf{d}_\mathsf{rm}$ at which a target can be reliably detected (i.e. with probability 1). Given the deterministic nature of the incoming target echo power, the metric readily follows as the solution of $\mathcal{S}(\mathsf{d}_\mathsf{rm}) = \theta$.

As far as comm operations are concerned, we assume a threshold-based decoding model. Accordingly, a packet sent over slot $m$ is successfully decoded if the received signal-to-interference ratio (SIR) exceeds a predefined value $\gamma$, i.e. with success probability $\mathsf{P}_\mathsf{s} = \mathbb{P}\{\mathsf{SIR} > \gamma\}$, where $\mathsf{SIR} = K d_c^{-\alpha}/\mathcal{I}_m$. The network performance will be evaluated in terms of throughput density $\mathcal{T}$, expressing the average number of packets successfully received per unit of time and area. Following the $q_c$-persistent slotted ALOHA under study,

$$\mathcal{T} = (1-\varepsilon) q_c \lambda \mathsf{P}_\mathsf{s}. \quad (3)$$

Unless otherwise specified, we assume $\alpha = 4$, $\varphi = \pi/6$, $\sigma = 10\,\mathrm{m}^2$, $M_r = 100$, $\mathsf{P}_\mathsf{f} = 0.1$, $\gamma = 5$ and $d_c = 5\,\mathrm{m}$.

## III. Radar Performance

As discussed in Sec. II, the detection threshold $\theta$ shall be set to guarantee a desired false alarm probability. Following (2), this requires the calculation of the joint distribution of the interference the typical radar receiver experiences over the $M_r - 1$ slots devoted to target detection. While conceptually simple, the task becomes mathematically intractable for practical values of $M_r$, due to the time-correlated nature of the interference that characterizes the setup under study. To overcome the issue and derive compact and insightful results we thus follow a different approach, tuning the detection threshold of a radar solely based on the disturbance coming from its nearest aligned interferer.[2] We can thus state

*Theorem 1:* Under the nearest interferer approximation, a desired $\mathsf{P}_\mathsf{f}$ is achieved by setting the detection threshold as

$$\theta = \frac{\varphi^4 \lambda^2 K}{16\pi^2 \ln^2\left(1 - \frac{\mathsf{P}_\mathsf{f}}{C(M_r, \varepsilon, q_c)}\right)} \quad (4)$$

where

$$C(M_r, \varepsilon, q_c) = 1 - \frac{\varepsilon}{M_r} - \sum_{i=M_r}^{M_r/\varepsilon - 1} \frac{\varepsilon}{M_r}(1 - q_c)^{N_i} \quad (5)$$

and $N_i = \min(M_r - 1, \frac{M_r}{\varepsilon} - i)$. The corresponding radar range evaluates to

$$\mathsf{d}_\mathsf{rm} = \left(\frac{4\pi\sigma}{\varphi^4 \lambda^2}\right)^{1/8} \left(-\ln\left(1 - \frac{\mathsf{P}_\mathsf{f}}{C(M_r, \varepsilon, q_c)}\right)\right)^{1/4} \quad (6)$$

*Proof:* Denote by $\mathbf{x}_1 \in \Phi$ the coordinates of the strongest potential interferer for the typical receiver (i.e. the closest node whose antenna pattern overlaps with the one of the typical receiver), and let $r_1 = \|\mathbf{x}_1\|$. By simple geometrical arguments combined with well-known stochastic geometry results [7], we get $F_{r_1}(x) := \mathbb{P}\{r_1 \leq x\} = 1 - \exp(-\lambda_a \pi x^2)$, where $\lambda_a = (\varphi/(2\pi))^2 \lambda$. Indicate now as $\mathsf{P}_\mathsf{n}(i)$ the probability that $\mathbf{x}_1$ does not perform any transmission (either in radar or comm mode) during the $M_r - 1$ slots spent by the typical receiver waiting for a target echo, conditioned on $\nu_1 = i$, $i \in \{0, 1, ..., M-1\}$. Recalling that $\nu_0 = 0$ (i.e. that the typical node sends its radar pulses at slots $nM$, $n \in \mathbb{N}$), $\mathsf{P}_\mathsf{n}(0) = 1$. Conversely, we clearly have $\mathsf{P}_\mathsf{n}(i) = 0, \forall i \in \{1, 2, ..., M_r - 1\}$. Consider instead the case $\nu_1 = M_r$ (reported in the second row of Fig. 1). In this situation, $\mathbf{x}_1$ does not interfere provided it sends no data packets over the previous $\min(M_r - 1, \frac{M_r}{\varepsilon} - M_r)$ slots it spent in comm mode. The event has probability $\mathsf{P}_\mathsf{n}(M_r) = (1 - q_c)^{\min(M_r - 1, \frac{M_r}{\varepsilon} - M_r)}$. Following a similar approach, $\mathsf{P}_\mathsf{n}(i)$ can be derived for any $i \in \{M_r+1, ..., M-1\}$, although not reported here due to space constraints. Recalling that $\mathbb{P}\{\nu_1 = i\} = 1/M \; \forall i$, the conditioning on the mark can readily be removed. We thus obtain the probability for $\mathbf{x}_1$ to actually interferere with the typical radar receiver as

$$1 - \mathsf{P}_\mathsf{n} = 1 - \frac{1}{M} - \sum_{i=M_r}^{M-1} \frac{1}{M}(1-q_c)^{N_i} \quad (7)$$

where $N_i = \min(M_r - 1, \frac{M_r}{\varepsilon} - i)$ expresses the number of slots $\mathbf{x}_1$ spent in comm mode in the period $\{1, ..., M_r - 1\}$ given that $\nu_1 = i$, $i \in \{M_r, ..., M-1\}$. A false alarm event is then induced by $\mathbf{x}_1$ with probability

$$\mathsf{P}_\mathsf{f} = (1 - \mathsf{P}_\mathsf{n}) \mathbb{P}\{K r_1^{-\alpha} > \theta\} \\ = C(M_r, \varepsilon, q_c) F_{r_1}\left((K/\theta)^{1/\alpha}\right) \quad (8)$$

where $C(M_r, \varepsilon, q_c)$, reported in (5), follows from (7) recalling that $M = M_r/\varepsilon$. Plugging the expression of $F_{r_1}(x)$ into (8), $\theta$ readily follows (4). Similarly, setting $\mathcal{S} = \theta$ and solving with respect to the target distance leads to $\mathsf{d}_\mathsf{rm}$ reported in (6). ∎

---

[1] We assume $M_r$ to be set so that the target range is limited by incoming power rather than by the unambiguous range.

[2] By nearest aligned interferer we mean the closest neighbour whose beam overlaps with the one of the typical receiver.

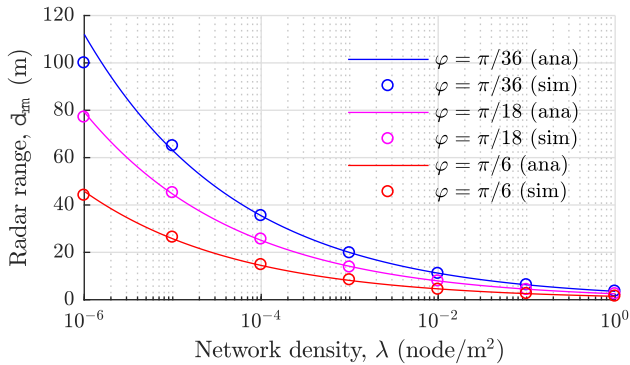

Fig. 2. Radar range $d_{rm}$ vs. $\lambda$. Lines show analytical results, while markers the outcome of simulation considering the aggregate interference in the network. $\varepsilon = 0.5$, $q_c = 0.5$. For simulations, $P_t = 10\,\text{dBm}$, $f = 60\,\text{GHz}$.

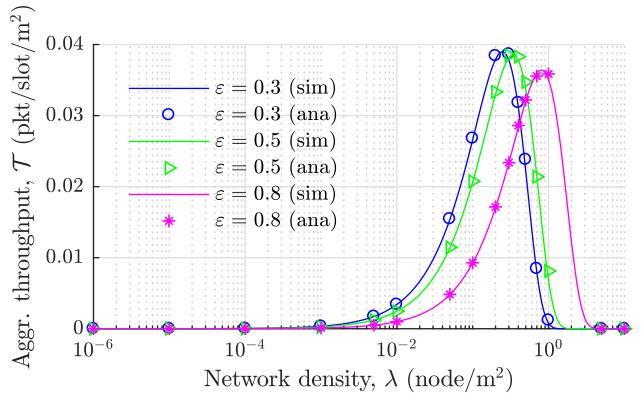

Fig. 3. Network throughput density $\mathcal{T}$ vs. $\lambda$, with $q_c = 0.5$. Lines show analytical results, while markers simulation ones considering the aggregate interference in the network. For simulations, $P_t = 10\,\text{dBm}$, $f = 60\,\text{GHz}$.

Theorem 1 offers interesting hints in terms of both radar design and performance evaluation, providing compact closed-form expressions that relate $\theta$ and $d_{rm}$ to the key system parameters. Notably, the detectable range does not depend on transmission power, operating frequency and antenna gain, and exhibits a small influence of the target radar cross section. The impact of network density and antenna directivity on $d_{rm}$ captured by (6) is instead reported in Fig. 2, assuming $\varepsilon = 0.5$ and $q_c = 0.5$. The plot clearly indicates the detrimental role of interference on the radar range, highlighting how practical detection distances can only be achieved in sparse topologies unless highly directional antennas are used.[3] To verify the accuracy of the strongest interferer approximation, dedicated network simulations were performed and the detection threshold was set to meet the target false alarm rate considering the *aggregate* interference. The outcome of this study (circle markers in Fig. 2) shows a very tight match with the analysis.

## IV. COMMUNICATION PERFORMANCE

To evaluate the network performance in terms of communication links, we start by computing the probability of successful packet delivery. Recalling (1), we readily get

$$\mathsf{P_s} = \mathbb{P}\{\mathcal{I} < K d_c^{-\alpha}/\gamma\} = F_\mathcal{I}\left(K d_c^{-\alpha}/\gamma\right) \quad (9)$$

where $F_\mathcal{I}(\cdot)$ is the cumulative distribution function of the aggregate interference at the typical node.[4] This, in turn, can be derived via the Laplace transform of $\mathcal{I}$. More specifically, let $\mathcal{I}^{(r)}$ and $\mathcal{I}^{(c)}$ indicate the aggregate interference coming from nodes that transmit a radar pulse and a comm packet over the observed slot, respectively, so that $\mathcal{I} = \mathcal{I}^{(r)} + \mathcal{I}^{(c)}$. As discussed in Sec. III, the corresponding PPPs $\Phi_r$ and $\Phi_c$ are independent. Thus, we have $\mathcal{L}_\mathcal{I}(s) = \mathcal{L}_{\mathcal{I}^{(r)}}(s)\mathcal{L}_{\mathcal{I}^{(c)}}(s)$, where $\mathcal{L}_X(s) := \mathbb{E}[e^{-sX}] = \int e^{-sx} f_X(x) dx$ indicates the Laplace transform of the r.v. $X$, and $f_X(x)$ is its probability density function (PDF). For the system model under consideration, simple linear transformations of a well-known stochastic geometry result [7] lead to

$$\mathcal{L}_\mathcal{I}(s) = \exp\left(-(\lambda_r + \lambda_c) K^{2/\alpha} \pi \, \Gamma(1 - 2/\alpha) \, s^{2/\alpha}\right)$$

where $\Gamma(x) = \int_0^\infty t^{x-1} e^t dt$ is the Gamma function. For the special case $\alpha = 4$ being studied, inverse Laplace transformation offers a closed form for the PDF of the aggregate interference, which leads after integration to

$$F_\mathcal{I}(x) = \text{erfc}\left(\pi^{3/2}(\lambda_r + \lambda_c)\sqrt{K/x}/2\right) \quad (10)$$

showing a Lévy distribution of parameter $\pi^3(\lambda_r + \lambda_c)^2 K/2$. Combining (3), (9) and (10), the network throughput density can eventually be expressed as

$$\mathcal{T} = (1-\varepsilon) q_c \lambda \, \text{erfc}\left(\frac{\lambda \varphi^2 d_c^2}{8\gamma^2 \pi^{1/2}} \left(\frac{\varepsilon}{M_r} + (1-\varepsilon) q_c\right)\right) \quad (11)$$

The derived closed-form expression (11) conveniently isolates the contribution of radar-to-comm and comm-to-comm interference through two additive terms in the argument of the complementary error function. A first glance at this interplay is offered in Fig.3, which shows the ALOHA-like trend of $\mathcal{T}$ against $\lambda$ for different values of $\varepsilon$ and setting $q_c = 0.5$. For the considered persistency, nodes access the shared channel more often when sending data packets compared to when operating in radar mode, as in the latter case a single transmission slot is followed by $M_r - 1$ idle time units. Thus, an increase in $\varepsilon$ reduces the overall level of interference, raising the packet delivery probability. On the other hand, the more the time devoted to radar detection, the less frequently data can be delivered, the lower the achievable per-node throughput. As a result of this tradeoff, larger values of $\varepsilon$ are more beneficial from the communication standpoint at higher network densities, while a reduction of the radar usage pays off more in sparser setups, as well exemplified by the plot. A direct comparison of Figs. 2 and 3 also highlights how densities at which practical radar ranges are achieved (e.g. $\lambda < 10^{-3}$) are likely not to be appealing in terms of comm throughput, suggesting a critical tradeoff that will be explored in the remainder of the letter.

---

[3] For very low $\lambda$, the radar performance would ultimately be dictated by the receiver sensitivity. These details are omitted for the sake of simplicity.

[4] Without risk of confusion, the index of the slot over which the comm packet is sent is omitted for readability.

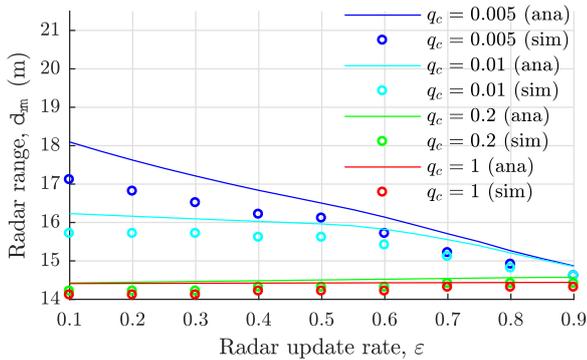 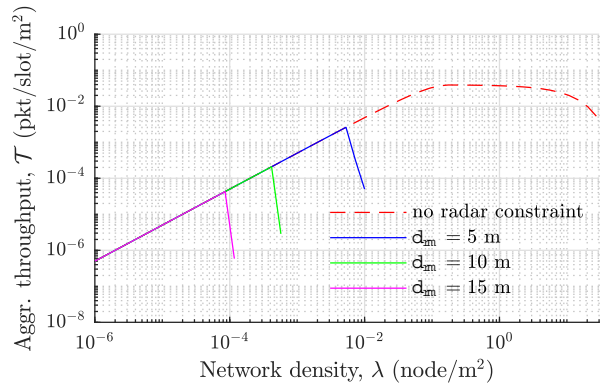

Fig. 4. Radar range $d_{rm}$ vs. $\varepsilon$. Lines show analytical results, while markers the simulation ones considering aggregate interference. $\lambda = 10^{-4}$.

Fig. 5. Maximum network throughput density $\mathcal{T}^*$ vs. $\lambda$. $\varepsilon = 0.5$.

## V. Radar-Communication Tradeoffs

To better understand the mutual influence of comm and radar activities, we report in Fig. 4 the detectable range obtained via (6) against the radar update rate $\varepsilon$, for different values of $q_c$ and setting $\lambda = 10^{-4}$. The approximated analytical results are verified by means of simulations that considered the aggregate level of interference in the network, showing a match within 5% in all conditions. Intuitively, a longer radar range can be achieved by reducing the comm persistency, as a result of the lower level of interference coming from data exchanges. For $q_c < 1/M_r$, i.e. if a device transmits more frequently when detecting a target than when operating in comm mode, the overall performance is dominated by radar-to-radar interference, and an increase in $\varepsilon$ has a detrimental effect on $d_{rm}$. The opposite trend emerges for $q_c > 1/M_r$, with a radar range that increases for higher $\varepsilon$, consistently with what discussed for the comm throughput in Fig. 3. Remarkably, in this case both persistency and radar update rate have a limited impact on $d_{rm}$, as instantiated by the curves for $q_c = 0.2$ and $q_c = 1$ in Fig. 4. Such a behaviour stems from the pivotal role played by the strongest interferer on radar performance. Indeed, a close inspection of (7) reveals that, for $q_c > 1/M_r$, the probability for the nearest aligned node to transmit at least once during the $M_r - 1$ slots spent by the typical receiver waiting for target echo rapidly approaches 1. Further increases in $q_c$ likely lead to multiple false alarms to be triggered during one detection cycle for the typical receiver. On the other hand, they do not significantly alter the statistics of the interference the node perceives over a single slot – which drive the tuning of the detection threshold – and eventually do not affect the overall performance. The reported results clearly highlight how challenging the coexistence of devices performing pulsed radar detection and data exchanges in an uncoordinated fashion over a shared channel can be for the former activity. Even at reasonably low densities (e.g. $\lambda = 10^{-4}$), and regardless of the fraction of time devoted to comm transmissions, interference has a disruptive effect on target detection unless a very low persistency for data exchanges – and thus a very low throughput – is imposed.

To further explore the tradeoff from the comm standpoint, we study how the persistency parameter shall be tuned so as to optimise the network throughput density. To this aim, we focus on a radar update rate $\varepsilon = 0.5$ and show in Fig. 5 the maximum throughput $\mathcal{T}^*$ achieved for any network density $\lambda$ by optimising over $q_c$, computed analytically from (11). The dashed curve indicates the achievable $\mathcal{T}^*$ disregarding the impact comm links have on radar performance. Conversely, solid lines are representative of an optimisation over $q_c$ subject to guaranteeing a minimum detectable range $d_{rm}$. In sparse topologies, the very low level of interference grants a high packet delivery rate and favours an aggressive use of the channel for data exchanges ($q_c = 1$ and linear growth of $\mathcal{T}^*$ with $\lambda$). When radar performance is not accounted for in tuning the persistency, the trend continues up to a critical density ($\lambda \simeq 0.1$), after which lower values of $q_c$ become more favourable to contain interference. Eventually, for very dense setups, the peak throughput plummets as expected, due to the uncoordinated nature of the considered channel access scheme. If, instead, a constraint on $d_{rm}$ is to be met, interference affecting radar detection becomes the most stringent factor, and data exchanges have to be deferred more often even in sparse networks. The severe cost undergone in terms of achievable throughput is evident in Fig. 5. Notably, after a certain density it becomes impossible to support $d_{rm}$ even in the absence of comm links, i.e. $q_c = 0$, stressing once more the sensitivity of radar operations within a field of interferers.

## References

[1] P. Kumari, J. Choi, N. Gonzlez-Prelcic, and R. W. Heath, "IEEE 802.11ad-Based Radar: An Approach to Joint Vehicular Communication-Radar System," *IEEE Trans. Veh. Technol.*, vol. 67, no. 4, pp. 3012–3027, April 2018.

[2] A. R. Chiriyath, B. Paul, G. M. Jacyna, and D. W. Bliss, "Inner Bounds on Performance of Radar and Communications Co-Existence," *IEEE Trans. Signal Process.*, vol. 64, no. 2, pp. 464–474, Jan 2016.

[3] M. Mehrnoush and S. Roy, "Coexistence of WLAN Network With Radar: Detection and Interference Mitigation," *IEEE Trans. on Cogn. Commun. Netw.*, vol. 3, no. 4, pp. 655–667, Dec 2017.

[4] A. Khawar, A. Abdelhadi, and C. Clancy, "Target Detection Performance of Spectrum Sharing MIMO Radars," *IEEE Sensors J.*, vol. 15, no. 9, pp. 4928–4940, Sept 2015.

[5] S. Shahi, D. Tuninetti, and N. Devroye, "On the Capacity of the AWGN Channel With Additive Radar Interference," *IEEE Trans. Commun.*, vol. 66, no. 2, pp. 629–643, Feb 2018.

[6] M. Braun, R. Tanbourgi, and F. K. Jondral, "Co-channel Interference Limitations of OFDM Communication-Radar Networks," *EURASIP Journal on Wireless Commun. and Netw.*, vol. 2013, no. 1, Aug 2013.

[7] M. Haenggi, *Stochastic Geometry for Wireless Networks.* Cambridge University Press, 2012.